# Outgassing rate comparison of seven geometrically similar vacuum chambers of different materials and heat treatments

Running title: Vacuum chamber comparison

Running Authors: Fedchak et al.

James A. Fedchak[1, a)], Julia K. Scherschligt[1], Sefer Avdiaj[2], Daniel S. Barker[1], Stephen P. Eckel[1], Ben Bowers[3], Scott O'Connell[3], and Perry Henderson[3]

[1]National Institute of Standards and Technology (NIST), 100 Bureau Dr., Gaithersburg, MD 20899, USA
[2]University of Prishtina, Department of Physics, Mother Teresa av 3, Prishtina 10000, Kosova
[3]Anderson Dahlen – Applied Vacuum Division, 6850 Sunwood Dr NW, Ramsey, MN 55303, USA

a) Electronic mail: james.fedchak@nist.gov

We have measured the water and hydrogen outgassing rates of seven vacuum chambers of identical geometry but constructed of different materials and heat treatments. Chambers of five different materials were tested: 304L, 316L, and 316LN stainless steels; titanium (ASTM grade 2); and 6061 aluminum. In addition, chambers constructed of 316L and 316LN stainless steel were subjected to a vacuum-fire process, where they were heated to approximately 950 °C for 24 hours while under vacuum. These latter two chambers are designated as 316L-XHV and 316LN-XHV. Because all the chambers were of identical geometry and made by the same manufacturer, a relative comparison of the outgassing rates among these chambers can be made. Water outgassing rates were measured as a function of time using the throughput technique. The water outgassing results for the 316L, 316LN, 316L-XHV, 316LN-XHV were all similar, but lower than those of 304L by a factor of 3 to 5 lower at $10^4$ s. The water outgassing results for Ti and Al chambers were closer to that of 304L, Ti being slightly lower. Hydrogen outgassing





rates were measured using the rate-of-rise method and performed after a low-temperature bake of 125 °C to 150 °C for a minimum of 72 hours. The Ti, Al, 316L-XHV, and 316LN-XHV chambers all have ultra-low specific outgassing rates below $1 \times 10^{-11}$ Pa L s$^{-1}$ cm$^{-2}$ and are a factor of 100 or better than the 304L chamber. The 304L, 316L, and 316LN chambers with no vacuum-fire heat treatment have larger hydrogen outgassing rates than the other chambers, with determined specific outgassing rates ranging between $4.0 \times 10^{-11}$ Pa L s$^{-1}$ cm$^{-2}$ and $8.0 \times 10^{-11}$ Pa L s$^{-1}$ cm$^{-2}$. We conclude that Ti, Al, 316L-XHV, and 316LN-XHV have hydrogen outgassing rates that make them excellent choices for ultra-high vacuum (UHV) and extreme-high vacuum (XHV) applications, the choice depending on cost and other material properties.

## I. INTRODUCTION

The ultimate pressure of a high vacuum, ultra-high vacuum (UHV) or extreme-high vacuum (XHV) system results from a balance between the system pumping speed and gas input. In a well-built system free from leaks or permeation, the gas input is still nonzero, because gas is emitted from materials that comprise the vacuum chamber and system components. This outgassing depends on the material and the way it produced, so its selection is of paramount importance to vacuum system design. A large body of literature exists for the outgassing rates of the common vacuum chamber materials stainless-steel,[1–8] aluminum,[9–11] and titanium;[12–16] the citations represent a small selection of the available literature. However, because of differences in materials, material





treatments, and measurement techniques, published outgassing rates often vary and questions arise as to the outgassing rate that can be practically achieved. Moreover, vacuum chambers have welds, seals, flanges, and variations in thickness and cleaning procedures that complicate the reproducibility of published outgassing rates. We present measured outgassing rates for seven vacuum chambers of identical geometry and construction, but of different vacuum materials. The seven chambers are made of titanium (grade 2); aluminum (6061); and five stainless steels of different types and heat-treatments: 304L, 316L, vacuum-fired 316L, 316LN electro-stag re-melt, and heat-treated 316LN electro-slag re-melt stainless-steel. All the vacuum chambers were fabricated, cleaned, and assembled by the same manufacturer (Anderson-Dahlen[17]). The vacuum-fire procedure (a 950 °C bake in vacuum for a minimum of 24 hours) was identical for the three vacuum fired chambers, designated here by -XHV. Outgassing rates were measured at NIST using SI-traceable measurement techniques. The same techniques and apparatus were used to measure the outgassing rates for all seven of the chambers. Therefore, by making relative comparisons of the outgassing rates among these seven chambers, we have sought to cancel the effects due to chamber construction and measurement method.

One goal of this study is to provide accurate data that can be used to select chamber materials for use in the vacuum. For UHV and XHV applications, vacuum chambers often require ultra-low specific outgassing rates of less than $1.0 \times 10^{-11}$ Pa L s$^{-1}$ cm$^{-2}$. Such a low outgassing rate cannot be achieved in stainless steel without heat treatment, generally at temperatures exceeding 400 °C.[3–5,18] Stainless steel is typically produced in an electo-arc process and, without additional refining processes, contains a high concentration of dissolved hydrogen. Aluminum contains far less





dissolved hydrogen than stainless steel, and titanium has been shown to have a diffusion barrier to hydrogen that leads to low outgassing rates.[12,13] We will demonstrate that aluminum, titanium, and heat-treated stainless steel all have excellent ultra-low outgassing rates and can be used to build vacuum chambers capable of obtaining UHV or XHV pressures (i.e., pressures below $1.0 \times 10^{-6}$ Pa).

Vacuum chambers undergo two distinct types of outgassing which need to be considered separately: that due to gases adsorbed on the surface (typically water), and that due to gases dissolved in the bulk (typically hydrogen). For vacuum systems constructed of the materials we consider here, outgassing of water from the chamber surface will dominate the outgassing rate during the first hours or days of evacuation from atmospheric pressure.[1] Water may be quickly desorbed from the surface by a low temperature bake in the nominal range of 100 °C to 250 °C. For chamber thicknesses exceeding 1 mm, bakes in this temperature range over many days will not significantly reduce the hydrogen outgassing that originates from gas dissolved in the bulk. After a low temperature bake, the outgassing products are predominantly hydrogen. We first determined the water outgassing rate as a function of time using a throughput method. Following a low-temperature bake, we then determined the hydrogen outgassing rate as a function of temperature using a rate-of-rise method. In both cases, the origin of the off-gassed products cannot be distinguished; i.e. the gas can desorb from the surface or originate from gas diffusing from the bulk material. The outgassing rate is the total throughput for the chamber and has units of Pa L s$^{-1}$. The outgassing rate per unit area, known as specific outgassing rate or outgassing flux, has units of Pa L s$^{-1}$ cm$^{-2}$ and can be considered a material property.





Because it is relatively easy to perform a low-temperature bake, water outgassing is not a relevant consideration compared to the hydrogen outgassing rate for many UHV and XHV applications. However, low-temperature bakes are not always possible, and, in some applications, long pump-down times are required to reduce pressure. For example, for some large vacuum systems, low-temperature bakes are prohibitively expensive, and often these systems may be baked once, at most, or may contain temperature-sensitive equipment and so cannot be baked at all. In these cases, the specific water outgassing rate or pump-down curve is an important consideration in vacuum design.

Two measurement techniques are used to determine the outgassing rates: the throughput method for determining water outgassing, and the rate-of-rise (RoR) method for determining the hydrogen outgassing. Details of these methods are discussed in section II. Briefly, in the RoR method the chamber is isolated from the vacuum pumps while under vacuum, and the outgassing rate is determined from the ensuing pressure rise in the chamber over time. In the throughput method, the chamber is evacuated through an orifice of known conductance while the chamber pressure is monitored, and the outgassing rate is determined from the calculated throughput, which is the product of the pressure and conductance. The throughput method can be used to determine a time-dependent outgassing rate and was therefore the preferred method for determining water outgassing because it tends to be strongly time-dependent during the time period of evacuation in unbaked vacuum systems. In the present experimental arrangement, this same method could not be used to determine the hydrogen outgassing rates because the resulting system pressure due to hydrogen outgassing, during the condition that the chamber is evacuated through the orifice, is too small to produce a significant signal in





the spinning rotor gauge (SRG) that is used to measure the chamber pressure. The RoR method has the advantage that the pressure rise can be recorded over long periods of time allowing an accumulation of pressure large enough to produce a strong signal, but is not as useful for outgassing rates that are strongly time-dependent or for gases that are strongly absorbed on the chamber surface, such as water.

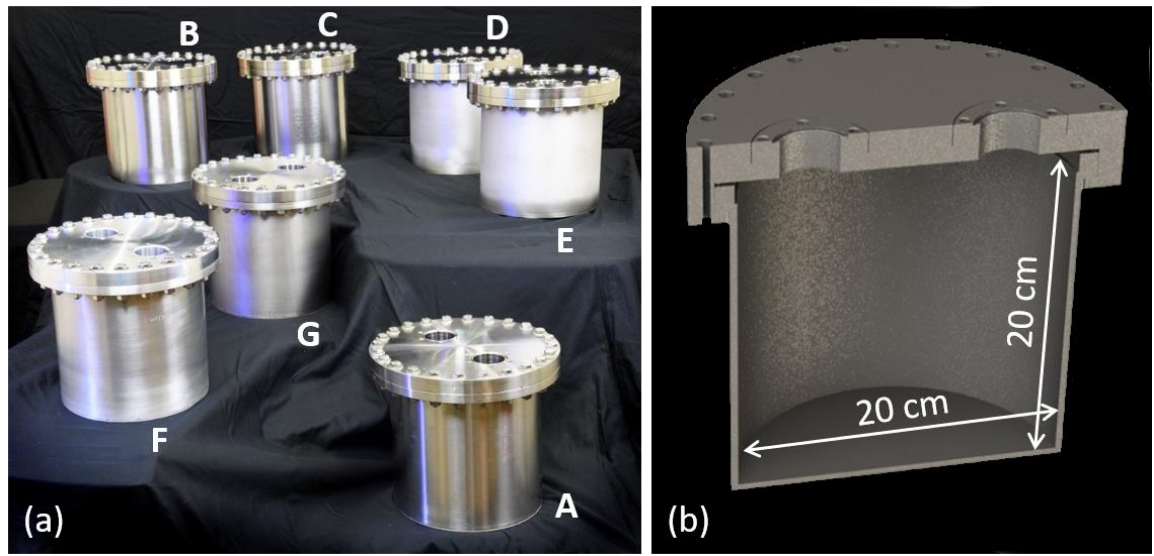

FIG. 1(a) Photograph of all sample chambers used in this study. These are (A) 304L, (B) 316L, (C) 316L-XHV, (D) 316LN, (E) 316LN-XHV, (F) Al, and (G) Ti. (b) Cross-section model view of the sample chamber.





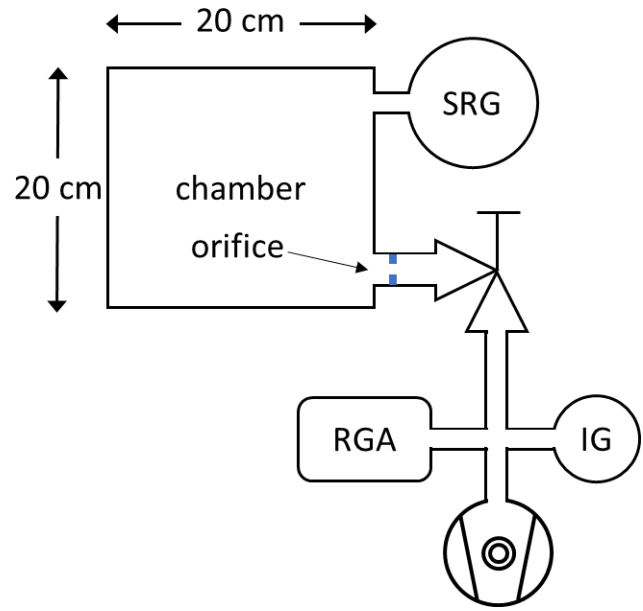

FIG. 2 Schematic of the apparatus used to determine the outgassing rates for the sample chambers. The water outgassing rates are determined using the throughput method and the hydrogen outgassing rates use the rate-of-rise method.

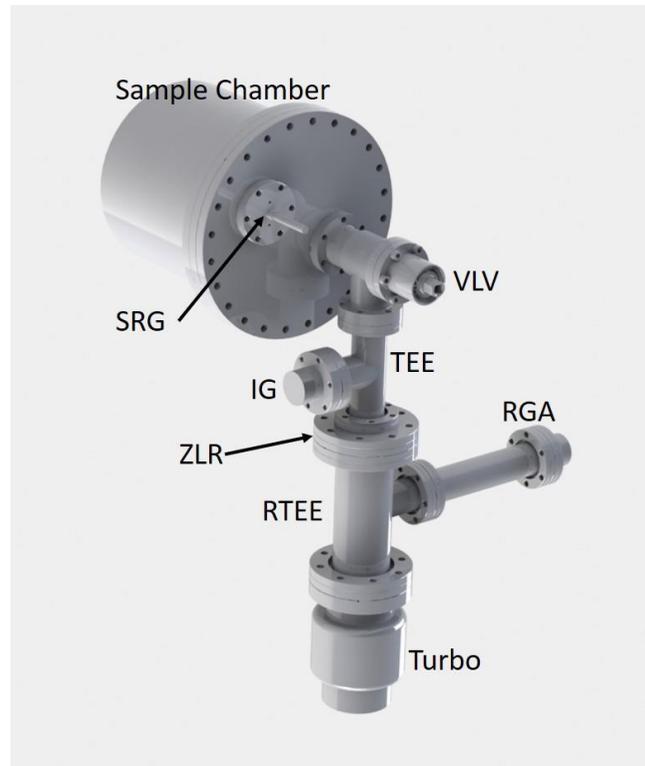





FIG. 3 Solid model showing approximate construction of the apparatus vacuum system. Not depicted is the fore-pump and fore-line that attaches to the back of the turbomolecular pump (Turbo).  SRG is the spinning rotor gauge; IG is an ionization gauge; VLV is the all-metal valve; TEE is a union tee; ZRL is a zero-length reducer flange; RTEE is a reducer tee; and RGA is a residual gas analyzer.

## II.  EXPERIMENTAL

### A.   Sample Chambers

Seven sample chambers were measured in this study; the material properties and designations are given in Table I and a picture of all the sample chambers is given in Fig. 1.  All chambers were fabricated by the same manufacturer to the same dimensions. Each chamber consists of three parts: a cylindrical body, a DN200 flange (throughout this paper the term DN*nnn* designates a knife-edge vacuum flange of nominal bore diameter *nnn* millimeters with dimensions specified in ISO 3669), and a sealing gasket. The DN200 flange and cylindrical body are made of the same materials, as given in Table 1. Flanges are cut from bar or plate stock, sheet metal is rolled for the main body of the chambers, and the chamber end is also cut from sheet metal.  Full chambers were then welded and cleaned using a UHV cleaning process in a designated clean room, which includes an aqueous rinse in an ultrasonic cleaning system followed by an acetone wipe down.  The cylindrical body is of 20 cm inner diameter and 20 cm interior length, open on one end and with 3 mm thick walls.  A knife-edge flange terminates one end of the cylinder and is sealed to a 13 mm thick DN200 flange using an OFHC (oxygen-free high thermal conductivity) copper gasket for all but the Al chamber, which used a gasket made of 1100-H14 series aluminum.  The thickness of this flange is thinner than a standard





DN200 flange to facilitate better degassing for the chambers that were vacuum fired. All interior surfaces were a standard machine finish of Ra 1.6 µm or less (Ra is the arithmetic average surface roughness). The volume of the sample chamber is 6.3 L with an interior surface area of approximately 2000 cm$^2$.

Two of the chambers, the 316L-XHV and 316LN-XHV where subjected to a vacuum fire process. The vacuum-fire process was performed by placing the raw materials used to fabricate the cylindrical body and DN200 flange in a vacuum furnace, where they were baked at approximately 950 °C for a minimum of 24 hours with the vacuum furnace pressure below $1 \times 10^{-2}$ Pa. After the vacuum-fire process, fabrication was performed on the raw materials to produce the cylindrical chamber body and flange, including cutting the knife-edge surfaces for the seals. The sealing gasket for the DN200 flange was not subjected to the vacuum-fire process. The DN200 flange was bolted to the chamber body at the factory for all seven of the chambers.

The DN200 flange has two symmetrically placed DN40 ports, each located 5.7 cm from the flange center and 11.4 cm from each other. An all-metal valve is attached to one port and to the other is attached a spinning rotor gauge (SRG)[19]. Note that the same valve and SRG were used for each chamber in turn. The interior surface of the SRG and the all-metal valve, up to the sealing surface, contribute to the outgassing rate during the rate-of-rise measurements; the combined surface area of the SRG plus valve represents less than 5 % of the total area of the sample chamber. Both the all-metal valve and SRG were baked at a temperature greater than 400 °C for two weeks to reduce the hydrogen outgassing rate.[18,20] The hydrogen outgassing rate of the combined SRG and all metal-





valve were measured in a separate experiment and subtracted from the chamber outgassing results.

      To use the throughput method for measuring the water outgassing rate, it is necessary to have a flow-constricting element with known conductance through which the outgassed water passes before it is evacuated. We achieve this by modifying a blank copper gasket (i.e., a solid 2 mm thick disk) to have a 6 mm diameter hole in the center. This blank gasket thus forms an orifice, which is installed between the chamber and the valve. The conductance of the orifice was determined from its dimensions to be 3.15 L s$^{-1}$ for $H_2O$ at 25 °C, with a $k = 2$ uncertainty (95 % confidence interval)[21] of less than 1 %. The orifice gasket must be replaced with a new one every time a sample chamber is installed; the above stated conductance uncertainty takes the dimensional variation into account. For the Al chamber, the orifice gasket was made of 1100-H4 aluminum to the same dimensions and uncertainty as stated above.

Table I. Description of all sample chambers used in this study. The material composition is taken from the certificates provided by the supplier.

| Designation | Material | Composition (percent) | Heat Treatment |
|---|---|---|---|
| Ti | titanium, ASTM grade 2 (unalloyed) | C, 0.007 to 0.012; N, $\leq 0.01$; Fe, 0.08 to 0.12; $O_2$, 0.1 to 0.14; other total, $\leq 0.17$; Ti, remainder | None |
| Al | 6061-T651 aluminum | Si, 0.4 to 0.9; Fe, 0.5 to 0.7; Cu, 0.15 to 0.4; Mn, 0.1 to 0.15; Mg 0.8 to 1.3; Cr, 0.18 to 0.35; Zn, 0.09 to 0.25; | None |

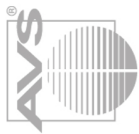





| | | | |
|---|---|---|---|
| | | Ti 0.02 to 0.15;<br>other total, 0.05 to 0.15;<br>Al, remainder of total | |
| 304L | 304L stainless steel | C, 0.017 to 0.027;<br>Si, 0.36 to 0.4;<br>Mn, 1.38 to 1.67;<br>P, 0.028 to 0.031;<br>S, 0.001 to 0.03;<br>Cr, 18.14 to 18.24;<br>Mo, 0.28 to 0.43;<br>Ni, 8.06 to 8.13;<br>N, 0.066 to 0.89;<br>Cu, 0.42 to 0.46;<br>Co, $\leq$ 0.121;<br>Fe, remainder | None |
| 316L | 316L stainless steel | C, 0.012 to 0.018;<br>Si, 0.38 to 0.5;<br>Mn, 1.18 to 1.69;<br>P, 0.03 to 0.031;<br>S, 0.001 to 0.03;<br>Cr, 16.54 to 16.7;<br>Mo, 2.01 to 2.08;<br>Ni, 10.0 to 10.02;<br>N, 0.04 to 0.057;<br>Cu, 0.11 to 0.36;<br>Co, $\leq$ 0.155;<br>Fe, remainder | None |
| 316L-XHV | 316L stainless steel | C, 0.012 to 0.018;<br>Si, 0.38 to 0.5;<br>Mn, 1.18 to 1.69;<br>P, 0.03 to 0.031;<br>S, 0.001 to 0.03;<br>Cr, 16.54 to 16.7;<br>Mo, 2.01 to 2.08;<br>Ni, 10.0 to 10.02;<br>N, 0.04 to 0.057;<br>Cu, 0.11 to 0.36;<br>Co, $\leq$ 0.155;<br>Fe, remainder | Vacuum-fired, 950 °C, $\geq$ 24 hours |
| 316LN | 316LN electroslag re-melt stainless steel | C, $\leq$ 0.012;<br>Si, 0.33 to 0.49;<br>Mn, 1.59 to 1.82;<br>P, 0.23 to 0.31;<br>S, 0.001 to 0.0018; | None |





| | | Cr, 17.16 to 17.38; Mo, 2.6; Ni, 12.24 to 13.55; N, 0.1636 to 0.1636; Co, ≤ 0.04; Fe, remainder | |
|---|---|---|---|
| 316LN-XHV | 316LN electroslag re-melt stainless steel | C, ≤ 0.012; Si, 0.33 to 0.49; Mn, 1.59 to 1.82; P, 0.23 to 0.31; S, 0.001 to 0.0018; Cr, 17.16 to 17.38; Mo, 2.6; Ni, 12.24 to 13.55; N, 0.1636 to 0.1636; Co, ≤ 0.04; Fe, remainder | Vacuum-fired, 950 °C, ≥ 24 hours |

## B. Outgassing apparatus

A schematic of the outgassing apparatus is given in Fig. 2 and a solid model showing the salient vacuum components is given in Fig. 3. The sample chamber is evacuated by a turbomolecular pump through an all-metal valve connected to the sample chamber; a dry mechanical pump backs the turbomolecular pump (not shown). An ionization gauge and a residual gas analyzer (RGA) are connected to the system between the turbomolecular pump and the all-metal valve. Both the ionization gauge and the RGA are used as diagnostics for the vacuum system and are not used to determine the outgassing rates. As described in the previous section, an orifice is between the all-metal valve and sample chamber, and an SRG is attached to the second port on sample chamber. The SRG is used to determine the outgassing rates; the ionization gauge is only used to monitor the background vacuum level. Several platinum-resistance thermometers (PRTs) are attached





to the chamber. The calibration of the PRTs are known to within an uncertainty of 50 mK. To determine the outgassing rates as a function of temperature, a temperature-controlled enclosure is placed over the sample chamber. The temperature difference across the chamber is less than 2 K in this arrangement. In general, the temperature was stable to within 1 K during the measurements.

Time-dependent water outgassing rates are determined using the throughput method. Prior to measuring water outgassing rates, the chambers are exposed to atmospheric conditions for at least several days, which is, presumably, long enough for the absorbed water to be in equilibrium with that in the atmosphere. A dry pump backing the turbomolecular pump (not shown in Fig. 2) evacuates the system up to the all-metal valve to a pressure $p < 10$ Pa while the sample chamber remains at atmospheric pressure. Evacuation of the chamber begins by opening the all-metal valve with the backing pump operating. The measurement system is then turned on as quickly as feasible. Within the first minute after opening the all-metal valve to begin evacuation, the turbomolecular pump is started, the SRG is suspended and operated, and data logging is initiated. This is defined as $t = 0$ s. Initially the SRG reading changes rapidly for two reasons: First, atmospheric gas is evacuated from the chamber in addition to water or other molecules desorbing from the chamber surfaces. Second, the SRG requires approximately 5 hours from initial suspension to come to temperature equilibrium with the chamber and produce a stable reading.[22] The rotor heats during the electromagnetic suspension causing the rotor temperature to rise; this heat slowly dissipates radiatively until the rotor temperature reaches equilibrium with the chamber, causing the rotor diameter and hence its moment of inertia to change.[22,23] We cut the first 30 min of data, after which the majority of the





gas in the chamber is due to desorption, and the rotor is in stable suspension and operating in the linear regime,[24,25] although the rotor signal is still changing as it comes to temperature equilibrium. For the SRG used here, this effect results in an error in the SRG background reading that is approximately 7 % low at $t_3 = 2.0 \times 10^3$ s, 2 % low at $t_4 = 1.0 \times 10^4$ s, and negligible at $t_5 = 1.0 \times 10^5$ s.

Pressure in the sample chamber $p$ is related to the gas throughput $q$ through the orifice by $q = (p - p_0)C$ where $p_0$ is the pressure downstream of the orifice with conductance $C$. The pressure ratio, defined as $R_p \equiv \dfrac{p}{p_0}$, is estimated to be $R_p = 6.0$ from the effective pumping speed $S$ downstream of the orifice and the conductance:

$R_p = \dfrac{S}{C} + 1$ . Table II gives the estimated conductance for each component used in the estimate of $S$. For the turbomolecular pump, the "conductance" is the pumping speed for water which we take to be the same as that for $N_2$. This seems to be a reasonable estimate,[26] and we note that a 30% change in the turbomolecular pump pumping speed only results in a 1% change in $Q_{H2O}$. With $Q = q/A$, where $A$ the interior surface area of the chamber assembly taken from its dimensions, the measurement equation for the specific outgassing rate of water using the throughput method becomes:

$$Q_{\text{H2O}}(t) = \left( \frac{R_p - 1}{R_p} \right) \frac{p(t)C_{\text{H2O}}}{A} . \qquad (1)$$

In eq.(1), $C_{\text{H2O}}$ is the $H_2O$ conductance of the orifice and $p(t)$ is the SRG pressure. A power-law of the form

$$Q_{\text{H2O}}(t) = Q_0 t^{-\alpha} \qquad (2)$$

is fit to the data.





TABLE II. Dimensions and conductance estimates used to estimate the pressure ratio $R_p$.

| Component | Length (mm) | Inner Diameter (mm) | Estimated Conductance (L/s) |
|---|---|---|---|
| VLV | --- | --- | 30.0[26] |
| TEE | 127.0 | 34.80 | 40.1 |
| ZLR | 12.7 | 34.80 | 401.5 |
| RTEE | 177.8 | 60.20 | 148.5 |
| Turbo | --- | --- | 75.0 |

The combined relative uncertainty in $Q_{H2O}$ is given by

$$u_{Q_{H2O}} = \sqrt{u_p^2 + u_{C_{H2O}}^2 + u_A^2 + \left(\frac{1}{R_p}\right)^2 u_{R_p}^2 + u_M^2 + u_{\text{type A}}^2}$$ ; all uncertainties reported here are given

with a coverage factor of $k = 2$ (95 % confidence interval). The uncertainty in the H$_2$O

conductance is estimated to be $u_{C_{H2O}} = 2$ % and $\frac{u_{R_p}}{R_p} \approx 3$ % is the uncertainty estimate due

the pressure ratio. The uncertainty in the pressure reading $u_p$ has three components: the

uncertainty of the absolute pressure reading, which is taken to be 6%;[22] the uncertainty

associated with the error in the reading caused by the rotor suspension as discussed

above; and the uncertainty due to the measured residual drag or background offset of the

SRG reading, which is estimated from the noise in the measurement and is negligible

compared to the other components.  Combined, these three components yield $u_p(t = 2 \times$

$10^3$ s) = 9 %, $u_p(t = 1 \times 10^4$ s) = 19 %, $u_p(t = 1 \times 10^5$ s) = 6 %.  Additionally, there is

uncertainty in the pressure reading associated with the assumption that the gas is entirely





water, this is accounted for in $u_M$. Both $p$ and $C_{H2O}$ depend on $\dfrac{1}{\sqrt{M}}$, therefore the

product $pC_{H2O}$ depends on $\dfrac{1}{M}$, where $M$ is the atomic mass of the gas sensed by the

SRG. After the first few minutes of evacuation, when free gases are removed, water and

hydrogen gas desorbing from the chamber surface constitute the majority of the gas left

in the chamber. Initially the largest fraction is water, but because water is more easily

removed, as the system is pumped for longer times hydrogen gas begins to dominate.

However, the water outgassing rate is determined by assuming all the outgassing

products are $H_2O$. Therefore, to estimate $u_M$, we assume 90 % of the gas is $H_2O$ at $t$

$=2000$ s, at $t = 10^4$ s 50 % is $H_2O$, and at $t = 10^5$ s only 10 % is $H_2O$. This yields $u_M(t = 2$

$\times 10^3$ s$) = 1$ %, $u_M(t = 1 \times 10^4$ s$) = 6$ %, and $u_M(t = 1 \times 10^5$ s$) = 10$ %. $u_A$ is the uncertainty

due to the chamber area and is negligible compared to the other components. $u_{\text{type A}}$ is the

type A uncertainty and is taken from the reproducibility of the water outgassing rate

discussed in Section III.b; it is the dominant uncertainty component and is approximately

$u_{\text{type A}} \approx 30$ %. Finally, to estimate $u_{Q_{H2O}}$, we take the root-mean-square average of the

combined uncertainties at $2 \times 10^3$ s, $t = 1 \times 10^4$ s, and $t = 1 \times 10^5$ s to obtain $u_{Q_{H2O}} = 33$ %.

Hydrogen outgassing rates are typically measured after the water outgassing

measurement is complete. Before beginning these measurements, the entire system is

baked to between 125 °C and 150 °C for a minimum of three days to remove most of the

remaining water. Afterwards, the sample chamber is allowed to re-equilibrate to

laboratory temperature (maintained below 26 °C) and the SRG is turned on. Hydrogen

outgassing rates are measured as a function of temperature. To reach temperatures above





laboratory temperature, the temperature-controlled box is placed around the sample chamber.

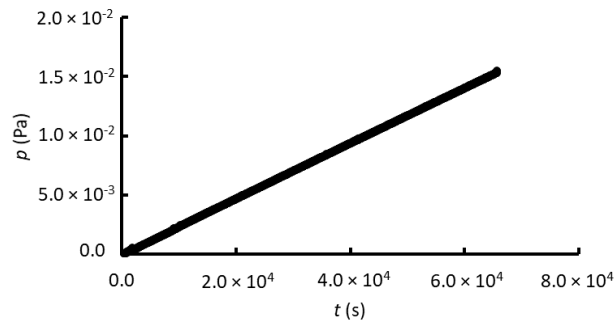

FIG. 4 An example of the rate-of-rise data used to determine the specific outgassing rate of the 304L chamber, showing the rise in pressure $p$ in the chamber as a function of time $t$.

Hydrogen outgassing is measured using a rate-of-rise technique. The vacuum level data produced by the SRG is continuously logged as a function of time throughout the entire process. The procedure begins with the all-metal valve open. The system temperature and pressure are monitored for stability. The all-metal valve is then closed and thus begins a hydrogen outgassing measurement. The pressure in the chamber rises with time, as shown in Fig. 4. The valve remains closed for at least 4 hours and up to several days. Opening the valve ends the rate-of-rise measurement; outgassing products are quickly evacuated, and the pressure burst may be observed on the residual gas analyzer (RGA).

A linear least-squares fit to the pressure vs. time data yields the slope $\frac{dp}{dt}$. The specific outgassing rate, or outgassing flux, is given by

$$Q_{H2} = \frac{V\dfrac{dp}{dt} - q_0}{A} \ . \qquad (3)$$





In eq. (3), $V$ is the volume of the sample chamber assembly (including the valve and SRG), $A$ is the macroscopic surface area of the sample chamber interior taken from its dimensions, and $q_0$ is the background outgassing rate due the SRG and valve determined in a separate rate-of-rise measurement with no sample chamber and the SRG directly connected to the valve.

The relative uncertainty of $Q_{H2}$ is given by $u_{Q_{H2}} = \sqrt{u_V^2 + u_p^2 + u_{q_0}^2 + u_A^2 + u_{\text{type A}}^2}$. The uncertainties associated with the chamber volume and area are $u_V$ and $u_A$ respectively, and are each estimated to be 4 %. The term $u_p$ is the uncertainty of the determined slope $\frac{dp}{dt}$. Following Sefa et al.[3], the major uncertainty contributor to $u_p$ is the SRG calibration factor for $H_2$ and $u_p$ is estimated to be 7 %. The uncertainty of the background outgassing is $u_{q_0}$, which has a $u_p$ component in addition to a component due the uncertainty in the volume of the SRG plus valve; it is estimated to be $u_{q_0} = 12$ %. The type-A uncertainty $u_{\text{type A}}$ is the uncertainty determined using statistical methods. The reproducibly of the measured $Q_{H2}$ is the dominant contributor to $u_{\text{type A}}$ and the combined uncertainty $u_{Q_{H2}}$. The conditions of measurement reproducibility are the ability to reproduce the measured outgassing rate after the chamber has been vented, exposed to the atmosphere for some time, placed back under vacuum and re-baked. For $Q_{H2} > 1.0 \times 10^{-11}$ Pa L s$^{-1}$ cm$^{-2}$, the uncertainty due to reproducibility is estimated to be 10 % from two measurements of the 316L chamber and is consistent with the reproducibility uncertainty of Sefa et al.[3] For $Q_{H2} < 1.0 \times 10^{-11}$ Pa L s$^{-2}$ the uncertainty due to reproducibility is 70 %, estimated from three measurements of the 316LN-XHV chamber and two





measurements of the 316L-XHV chamber . Real changes in the surface conditions may explain much of the irreproducibility for $Q_{H2} < 1.0 \times 10^{-11}$ Pa L s$^{-1}$ cm$^{-2}$; however, measurement noise is also a significant contributor. The background $q_0$ and SRG noise limit the measurable specific outgassing rate to above $2 \times 10^{-13}$ Pa L s$^{-1}$ cm$^{-2}$ for data collection times on the order of a day. In addition, the fundamental background signal of the SRG (i.e. the residual drag) can have a frequency dependence that must be considered and can be another significant contributor to the type A uncertainty for low signals. Finally, $u_{Q_{H2}} = 24$ % for $Q_{H2} > 1.0 \times 10^{-11}$ Pa L s$^{-1}$, and $u_{Q_{H2}} = 72$ % for $Q_{H2} < 1.0 \times 10^{-11}$ Pa L s$^{-1}$.







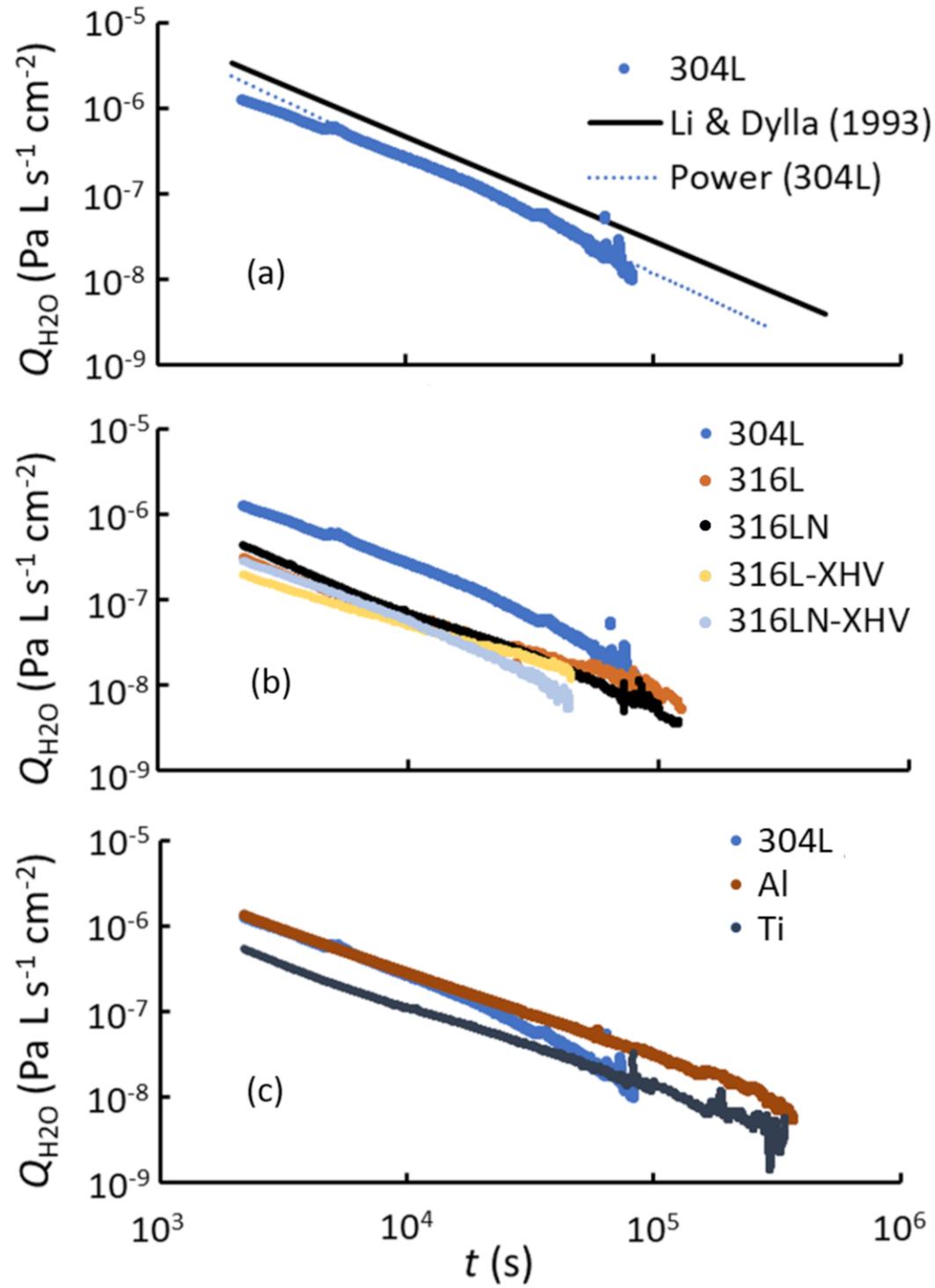

FIG. 5 The measured specific outgassing rates for water, $Q_{H2O}$, for all seven sample chambers. (a) The present results for 304L compared to the results of Li and Dylla[1] for





chambers vented with air; (b) the present results for 304L compared to those of 316L, 316LN, 316L-XHV, and 316LN-XHV; (c) The present results for 304L compared to those of Ti and Al.

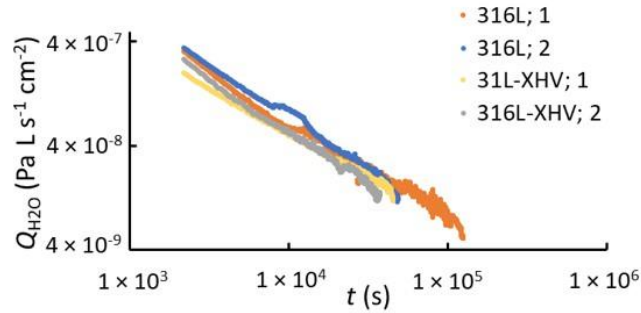

FIG. 6 Repeat measurements of the specific outgassing rates for water, $Q_{H2O}$, for 316L and 316L-XHV. The chambers were removed from the apparatus and left in the ambient atmosphere between repeat measurements. 316L;1 is the same data presented in Fig. 5(b); 316L;2 is a repeat measurement taken more than 1 year later. 316L-XHV;1 is the same data as in Fig. 5(b); 316L-XHV;2 is a repeat measurement take approximately 9 months later.

## III. Results and Discussion

### A.   Water Outgassing

Measured water outgassing rates are shown in Figs. 5 and 6. We include the water outgassing results from Li and Dylla[1] for a 304L chamber exposed to ambient air in Fig. 5(a) to facilitate a relative comparison among the present results to those benchmark results.  The dotted line in Fig. 5(a) represents the power law fit of eq. (3) to the 304L data. The time span of the data varies among the measurement runs.  Data collection and analysis is terminated due to noise limits or other practical laboratory concerns such as changing environmental conditions.  Results of the power law fit for all seven chambers are summarized in Table III, along with the specific outgassing rates at three different times normalized to the those of the 304L chamber. Here the three times are $t_3 = 2 \times 10^3$ s, $t_4 = 10^4$ s, and $t_5 = 10^5$ s. Present results for the 304L chamber compare fairly well to the results of Li and Dylla; our results for $Q_{H2O}(t_3)$ are about 40 % lower than those of Li and Dylla and are a factor of 2.3 lower at $t = t_5$. Results for the 316 series chambers—316L, 316LN, 316L-XHV, and 316LN-XHV—all display lower initial outgassing rates than the 304L chamber. This is shown in Fig. 5(b). At $t = 10^5$ s, the 316L and 316L-XHV chambers have between 30 % and 50% lower outgassing rates than does the 304L chamber. Similarly, at $t = 10^5$ s, the 316LN and 316LN-XHV chambers have between 60 % and 80 % lower outgassing rates than does the 304L chamber. The time-dependence in $Q_{H20}$ for the 316LN and 316LN-XHV chambers are similar to that for the 304L chamber, as demonstrated by $\alpha$ in Table III, but the 316L and 316L-XHV chamber seem to lose water at a slower rate.

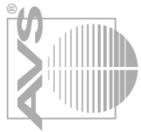





Table III. Fit parameters $Q_0$ and $\alpha$ from Eq. (2). Given the fit parameters, the specific outgassing rate for water $Q_{H2O}(t)$ is calculated at three times: $t_3 = 2 \times 10^3$ s, $t_4 = 10^4$ s, and $t_5 = 10^5$ s. The $Q_{H2O}(t)$ at these three times are also normalized to those of the 304L chamber. The uncertainty of $Q_{H2O}(t)$ is 33%.

| Chamber | $Q_0$ | $\alpha$ | $Q_{H2O}(t)$ (Pa L s$^{-1}$ cm$^{-2}$) | | | $Q_{H2O}(t) / Q_{H2O}(t;304L)$ | | |
|---|---|---|---|---|---|---|---|---|
| | | | $t_3$ | $t_4$ | $t_5$ | $t_3$ | $t_4$ | $t_5$ |
| 304L | $7.0 \times 10^{-2}$ | 1.3 | $2.4 \times 10^{-6}$ | $2.7 \times 10^{-7}$ | $1.2 \times 10^{-8}$ | 1 | 1 | 1 |
| 316L; 1 | $2.5 \times 10^{-4}$ | 0.89 | $2.8 \times 10^{-7}$ | $6.6 \times 10^{-8}$ | $8.5 \times 10^{-9}$ | 0.1 | 0.2 | 0.7 |
| 316L; 2 | $1.2 \times 10^{-3}$ | 1.0 | $4.1 \times 10^{-7}$ | $7.7 \times 10^{-8}$ | $6.9 \times 10^{-9}$ | 0.2 | 0.3 | 0.6 |
| 316L-XHV; 1 | $1.3 \times 10^{-4}$ | 0.83 | $2.0 \times 10^{-7}$ | $5.2 \times 10^{-8}$ | $7.7 \times 10^{-9}$ | 0.1 | 0.2 | 0.6 |
| 316L-XHV; 2 | $5.3 \times 10^{-4}$ | 0.99 | $2.6 \times 10^{-7}$ | $5.3 \times 10^{-8}$ | $5.4 \times 10^{-9}$ | 0.1 | 0.2 | 0.5 |
| 316LN | $4.1 \times 10^{-3}$ | 1.2 | $5.2 \times 10^{-7}$ | $7.8 \times 10^{-8}$ | $5.2 \times 10^{-9}$ | 0.2 | 0.3 | 0.4 |
| 316LN-XHV | $7.4 \times 10^{-3}$ | 1.3 | $4.2 \times 10^{-7}$ | $5.4 \times 10^{-8}$ | $2.8 \times 10^{-9}$ | 0.2 | 0.2 | 0.2 |
| Al | $4.2 \times 10^{-3}$ | 1.0 | $1.6 \times 10^{-6}$ | $3.2 \times 10^{-7}$ | $3.0 \times 10^{-8}$ | 0.7 | 1.2 | 2.5 |
| Ti | $1.3 \times 10^{-3}$ | 1.0 | $6.4 \times 10^{-7}$ | $1.3 \times 10^{-7}$ | $1.3 \times 10^{-8}$ | 0.3 | 0.5 | 1.1 |

Vacuum-firing does not seem to significantly affect the water outgassing. Neither the Al chamber nor the Ti chamber demonstrate significant practical advantage over 304L in terms of the outgassing rate at $t = 10^5$ s: Ti is about 10 % higher, which is within the measurement uncertainty, and the Al chamber is a factor of 2.5 higher.

Finally, the reproducibility is demonstrated in Fig. 6. Repeat measurements of $Q_{H2O}$, for 316L and 316L-XHV are shown. The chambers were removed from the apparatus and left in the ambient atmosphere between repeat measurements. The two data sets on the 316L chamber were collected more than 1 year apart, and two data sets on the 316L-XHV chamber were collected approximately 9 months apart. We point out that our exposure times were not well controlled in that they vary from several days to many months at a relative humidity ranging from 45 % to 55 %. Although water outgassing rates of metals are known to depend on the degree of exposure, given enough time the amount of water on the surface and absorbed in the near surface will eventually come to an equilibrium with that in the atmosphere. Indeed, the data in Li and Dylla (1994)[27] shows that for a stainless steel surface at 310 K, for example, the outgassing rate increases by a factor of 1.7 as the exposure time is increased from 1 min to 60 min, but only increases by a factor of 1.1 as the exposure time is increased from 60 min to 500 min.  Thus, we do not expect the water outgassing rate to significantly change after a few days of exposure.  A small "bump" observed in the repeated 316L data is due to a rapid temperature change in the laboratory which caused the SRG signal to change proportionally to the time derivative of the temperature during the event.[22,23] The reproducibility given in section II.B. is taken from data shown Fig. 6 and in Table III.

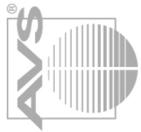

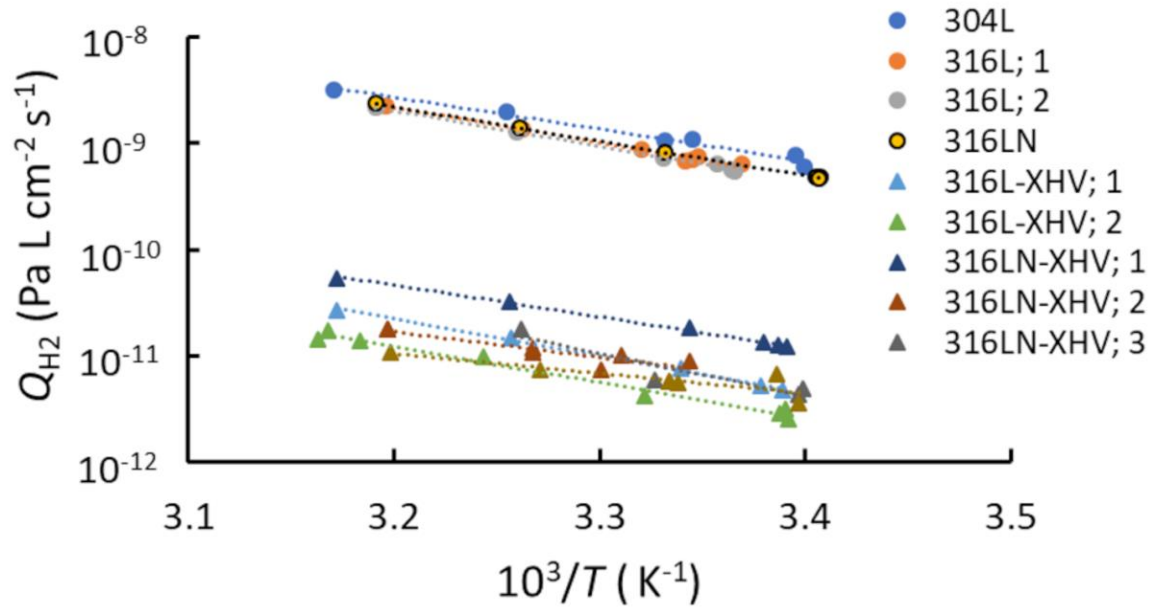

FIG. 7 Arrhenius plots of the measured specific outgassing rate for H2 $Q_{H2}$ for six chambers.





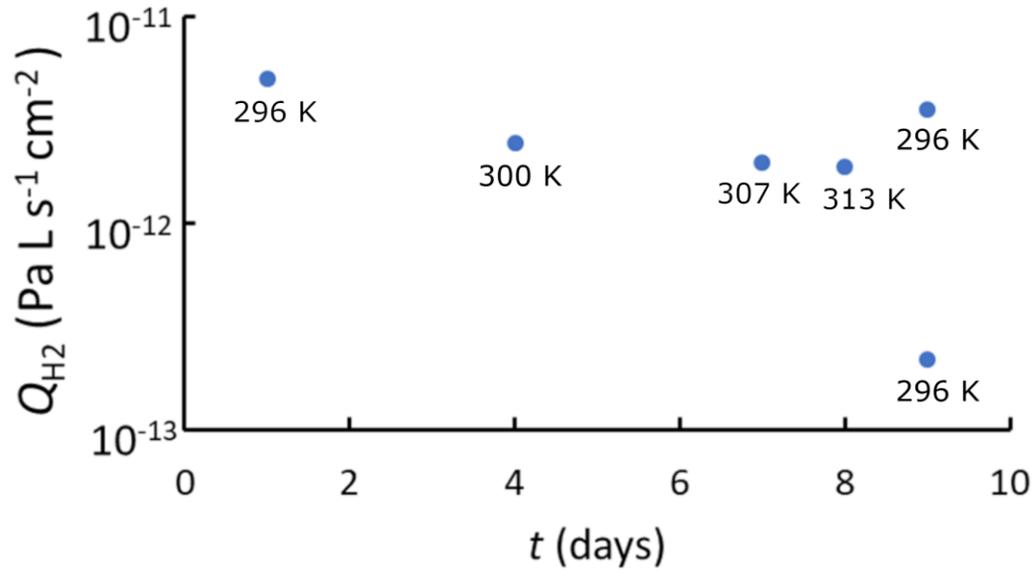

FIG. 8 Plots of the measured specific outgassing rate for H₂ $Q_{H2}$ for Ti as a function of relative time in days.

## B.   *Hydrogen Outgassing*

Fig. 7 shows measured temperature dependent hydrogen specific outgassing rates plotted on Arrhenius plots. In the limit of diffusion-limited outgassing, hydrogen outgassing originating from the bulk of the material may be described by

$$\log Q_{H2} = \log A_0 - \frac{E_D}{k_B T}, \qquad (4)$$

where $E_D$ is an activation energy for hydrogen diffusion and $A_0$ is related to the initial hydrogen concentration in the bulk of the material. Linear fits to the log of the outgassing data are displayed on all plots in Fig. 7. A summary of the $E_D$ and specific outgassing rates determined from the fit at 25 °C is given in Table IV. In cases where multiple data sets were taken, the average of all $Q_{H2}$ determined at 25 °C is given.  The present $E_D$





results for 304L and 316L are within 5 % and 20 %, respectively, of the benchmark results of Grant et al.[28,29] For Ti, the specific outgassing rate given in Table IV is the average of all Ti data. The temperature dependence of the outgassing results for the Ti chamber did not follow the Arrhenius relationship. These are plotted as a function of time in Fig. 8, where $t = 0$ is chosen as the time when the chamber bake was completed. The temperature of the chamber during the outgassing measurement is indicated by a label next to the data point. The general trend for the Ti chamber was that the outgassing rate decreased over time. This is not entirely surprising considering the studies done by Takeda and coworkers.[12,13] They demonstrate that the hydrogen concentration in Ti is largest in the boundary between the Ti bulk and oxide layer on surface, suggesting strong hydrogen traps in that region, but a much lower concentration of hydrogen in the surface oxide layer, suggesting a faster diffusion rate for hydrogen in the surface layer. These authors found that the oxide layer in Ti had much lower concentrations of hydrogen than the oxide layer in stainless steel. We speculate that in our experimental arrangement, hydrogen depletes from the Ti surface layer over a few days, reducing the outgassing such that it reaches the noise floor of the measurement. In Table IV, the specific outgassing rate that we give for Ti at 25 °C is the average of all the Ti measurements.

TABLE IV. Specific hydrogen outgassing rates $Q_{H2}$ at 298.15 K. For Ti, $Q_{H2}$ is taken as the average of all the Ti data. For the others, $Q_{H2}$ is determined from the fitting parameters $A_0$ and $E_D$ from Eq. (4). The uncertainty is $u_{Q_{H2}} = 24$ % for the 316L, 316LN, and 304L chambers, and $u_{Q_{H2}} = 72$ % for all others.

| Chamber | Activation Energy $E_D$ | | $Q_{H2}(T = 298.15 \text{ K})$ | Relative improvement factor |
|---|---|---|---|---|
| | eV | K | Pa L s$^{-1}$ cm$^{-2}$ | |
| Ti | | | $2.5 \times 10^{-12}$ | 377 |
| Al | 0.37 | 4250 | $5.5 \times 10^{-12}$ | 172 |
| 316L-XHV | 0.68 | 8080 | $5.1 \times 10^{-12}$ | 184 |







| 316LN-XHV | 0.53 | 6890 | $9.9 \times 10^{-12}$ | 95 |
| 316L | 0.66 | 7580 | $6.5 \times 10^{-10}$ | 1.5 |
| 316LN | 0.64 | 7450 | $7.0 \times 10^{-10}$ | 1.3 |
| 304L | 0.59 | 6880 | $9.4 \times 10^{-10}$ | 1.0 |

As can be seen from Table IV, ultra-low specific outgassing rates of $Q_{H2} < 1.0 \times 10^{-11}$ Pa L s$^{-1}$ cm$^{-2}$ are obtained from Al, Ti, 316L-XHV, and 316LN-XHV for $T = 298.15$ K. As previously discussed, the repeatability of the outgassing measurements for $Q_{H2} < 1 \times 10^{-11}$ Pa L s$^{-1}$ cm$^{-2}$ is roughly 50 %; therefore, results for the Al and the vacuum-fired stainless-steel chambers are equivalent to within the measurement uncertainty. For the Ti chamber, it is possible that there is re-absorption of the hydrogen on the Ti surface, and therefore the true outgassing rate may be larger than that measured using a rate-of-rise technique. Nevertheless, our results show that Ti is an excellent choice for many XHV applications. Table IV also give the outgassing improvement relative to 304L stainless steel. We chose 304L as a benchmark because it is one of the most commonly used materials for vacuum chambers in the United States and, as the present study shows, produces a similar hydrogen outgassing rate as 316L or 316LN with no heat treatment. We see no improvement in the hydrogen outgassing between the chambers constructed of 316LN, produced by electroslag re-melt, over the chambers constructed of 316L. This is significant because 316LN tends to be a more expensive material than 316L or 304L, and therefore may not be the most economical choice unless the application requires materials with low magnetic susceptibility, for example. The present results for Ti are in accord with previous outgassing measurements for Ti;[12,14] similarly, the present results for vacuum-fired 316L are comparable to or lower than outgassing measurement for vacuum-fired 316L found in the literature.[3–5,30]





# IV. SUMMARY AND CONCLUSIONS

We have compared the specific outgassing rates $Q_{H2}$ for seven chambers of identical geometry but of different materials and heat treatments:  Al, Ti, 304L, 316L, 316LN, 316L-XHV, and 316LN-XHV.   The chambers 316L-XHV and 316LN-XHV were vacuum fired at 950 °C for greater than 24 hours by Anderson-Dahlen.[17]  Hydrogen outgassing rates were determined after a low-temperature bake between 100 °C and 150 °C for a minimum of 3 days.  We found that Al, Ti, 316L-XHV, and 316LN-XHV all produced for $Q_{H2} < 1\times10^{-11}$ Pa L s$^{-1}$ cm$^{-2}$ and are excellent choices for UHV or XHV applications. Al, Ti, 316L-XHV, and 316LN-XHV show an improvement in outgassing of a factor of about 100 over 304L. Ti demonstrated the lowest $H_2$ outgassing, roughly a factor of 300 improvement over 304L.  It is possible that some re-adsorption of outgassed hydrogen occurs on the Ti surface. We conclude that Al, Ti, 316L-XHV, and 316LN-XHV are all excellent choices for UHV or XHV applications. Material cost and properties then become the more important consideration in choosing among these materials. For example, although 304L generally costs less than 316L, 316L contains molybdenum and is more corrosion resistant than 304L and is typically used for vacuum-firing (as in this study) because it is regarded as more resistant to softening during the firing process. All the stainless steels tested have excellent structural and mechanical properties (machinability, weldability, etc.), but 316LN is the least magnetic and is used in applications that require non-magnetic steel, even though this alloy tends to be expensive.  Al has the least expensive material cost compared to the other chambers





studied here, and has an excellent strength-weight ratio, but practical chambers made of Al can often be more expensive than those of stainless steel because of the difficulties of welding Al. Similarly, Ti tends to be an expensive material, and is more difficult to work with than stainless steel.

Water outgassing rates as a function of pump-down time where also determined for the seven chambers. The Al, Ti, and 304L chambers had similar water outgassing rates. The 316 stainless steels, 316L, 316L-XHV, 316LN, and 316LN-XHV, all started with lower water outgassing rates, about ten times lower than 304L at $2 \times 10^3$ s, but the 316L and 316L-XHV do not show a significant improvement over the 304L chamber at $10^5$ s. Interestingly, the 316LN and 316LN-XHV maintain a similar improvement in the water outgassing rate as a function of time to the 304L chamber, although the overall rate is about 10 times lower.

We have continued these studies to include other materials such as structural steel and will publish those additional results in the near future.

## ACKNOWLEDGMENTS


Many thanks to Joshua Levy for his work helping to put the system together, and to Toby Herman for his generosity in lending us lab space.

The data that supports the findings of this study are available within the article.



[1] M. Li and H.F. Dylla, J. Vac. Sci. Technol. A Vacuum, Surfaces, Film. **11**, 1702 (1993).

[2] K. Battes, C. Day, and V. Hauer, J. Vac. Sci. Technol. A J. Vac. Sci. Technol. A J. Vac. Sci. Technol. A **33**, 21603 (2015).

[3] M. Sefa, J.A. Fedchak, and J. Scherschligt, J. Vac. Sci. Technol. A Vacuum, Surfaces, Film. **35**, 041601 (2017).







[4] M.A.A. Mamun, A.A. Elmustafa, M.L. Stutzman, P.A. Adderley, and M. Poelker, J. Vac. Sci. Technol. A Vacuum, Surfaces, Film. **32**, 021604 (2014).

[5] C.D. Park, S.M. Chung, X. Liu, and Y. Li, J. Vac. Sci. Technol. A Vacuum, Surfaces, Film. **26**, 1166 (2008).

[6] Y.T. Sasaki, J. Vac. Sci. Technol. A Vacuum, Surfaces, Film. **25**, 1309 (2007).

[7] Y. Ishikawa and V. Nemanič, Vacuum **69**, 501 (2003).

[8] S. Avdiaj and B. Erjavec, Mater. Tehnol. **46**, 161 (2012).

[9] J.R. Young, J. Vac. Sci. Technol. **6**, 398 (1969).

[10] J.R. Chen, J.R. Huang, G.Y. Hsiung, T.Y. Wu, and Y.C. Liu, J. Vac. Sci. Technol. A Vacuum, Surfaces, Film. **12**, 1750 (1994).

[11] R. Dobrozemsky, S. Menhart, and K. Buchtela, J. Vac. Sci. Technol. A **25**, 551 (2007).

[12] M. Takeda, H. Kurisu, S. Yamamoto, H. Nakagawa, and K. Ishizawa, Appl. Surf. Sci. **258**, 1405 (2011).

[13] M. Takeda, H. Kurisu, S. Yamamoto, and H. Nakagawa, Vacuum **84**, 352 (2009).

[14] H. Kurisu, K. Ishizawa, S. Yamamoto, M. Hesaka, and Y. Saito, J. Phys. Conf. Ser. **100**, (2008).

[15] K. Ishizawa, H. Kurisu, S. Yamamoto, T. Nomura, and N. Murashige, J. Phys. Conf. Ser. **100**, (2008).

[16] M. Minato and Y. Itoh, J. Vac. Sci. Technol. A Vacuum, Surfaces, Film. **13**, 540 (1995).


[17] Certain commercial equipment, instruments, or materials are identified in this paper in order to specify the experimental procedure adequately. Such identification is not intended to imply recommendation or endorsement by NIST, nor is it intended to imply





that the materials or equipment identified are necessarily the best available for the purpose.


[18] J.A. Fedchak, J. Scherschligt, D. Barker, S. Eckel, A.P. Farrell, and M. Sefa, J. Vac. Sci. Technol. A Vacuum, Surfaces, Film. **36**, 023201 (2018).

[19] J.K. Fremerey, J. Vac. Sci. Technol. A Vacuum, Surfaces, Film. **3**, 1715 (1985).

[20] M. Sefa, J.A. Fedchak, and J. Scherschligt, J. Vac. Sci. Technol. A Vacuum, Surfaces, Film. **35**, 041601 (2017).

[21] B.N. Taylor, *Guidelines for Evaluating and Expressing the Uncertainty of NIST Measurement Results* (Gaithersburg, MD, 1994).

[22] S. Dittmann, B.E. Lindenau, and C.R. Tilford, J. Vac. Sci. Technol. A Vacuum, Surfaces, Film. **7**, 3356 (1989).

[23] J.A. Fedchak, K. Arai, K. Jousten, J. Setina, and H. Yoshida, Measurement **66**, 176 (2015).

[24] R.F. Berg and J.A. Fedchak, *NIST Calibration Services for Spinning Rotor Gauge Calibrations* (Gaithersburg, MD, 2015).

[25] J. Šetina, J. Vac. Sci. Technol. A **17**, 2086 (1999).

[26] K. Jousten, editor , *Handbook of Vacuum Technology* (WILEY-VCH Verlag GmbH & Co. KGaA, Weinheim, 2008).

[27] M. Li and H.F. Dylla, J. Vac. Sci. Technol. A Vacuum, Surfaces, Film. **12**, 1772 (1994).

[28] D.. Grant, D.. Cummings, and D.. Blackburn, J. Nucl. Mater. **149**, 180 (1987).

[29] D.. Grant, D.. Cummings, and D.. Blackburn, J. Nucl. Mater. **152**, 139 (1988).

[30] K. Jousten, Vacuum **49**, 359 (1998).






TABLES

TABLE I. Description of all sample chambers used in this study.

| Designation | Material | Heat Treatment |
|---|---|---|
| Ti | titanium, ASTM grade 2 (unalloyed) | None |
| Al | 6061 aluminum | None |
| 304L | 304L stainless steel | None |
| 316L | 316L stainless steel | None |
| 316L-XHV | 316L stainless steel | Vacuum-fired, 950 °C, $\geq$ 24 hours |
| 316LN | 316LN electroslag re-melt stainless steel | None |
| 316LN-XHV | 316LN electroslag re-melt stainless steel | Vacuum-fired, 950 °C, $\geq$ 24 hours |

TABLE II. Dimensions and conductance estimates used to estimate the pressure ratio $R_p$.

| Component | Length (mm) | Inner Diameter (mm) | Estimated Conductance (L/s) |
|---|---|---|---|
| VLV | --- | --- | 30.0[26] |
| TEE | 127.0 | 34.80 | 40.1 |
| ZLR | 12.7 | 34.80 | 401.5 |
| RTEE | 177.8 | 60.20 | 148.5 |
| Turbo | --- | --- | 75.0 |





Table III. Fit parameters $Q_0$ and $\alpha$ from Eq. (2). Given the fit parameters, the specific outgassing rate for water $Q_{H2O}(t)$ is calculated at three times: $t_3 = 2 \times 10^3$ s, $t_4 = 10^4$ s, and $t_5 = 10^5$ s. The $Q_{H2O}(t)$ at these three times are also normalized to those of the 304L

| | | | $Q_{H2O}(t)$ | | | $Q_{H2O}(t) \big/ Q_{H2O}(t; 304L)$ | | |
| | | | (Pa L s$^{-1}$ cm$^{-2}$) | | | | | |
| Chamber | $Q_0$ | $\alpha$ | $t_3$ | $t_4$ | $t_5$ | $t_3$ | $t_4$ | $t_5$ |
|---|---|---|---|---|---|---|---|---|
| 304L | 7.0E-02 | 1.3 | 2.4E-06 | 2.7E-07 | 1.2E-08 | 1 | 1 | 1 |
| 316L; 1 | 2.5E-04 | 0.89 | 2.8E-07 | 6.6E-08 | 8.5E-09 | 0.1 | 0.2 | 0.7 |
| 316L; 2 | 1.2E-03 | 1.0 | 4.1E-07 | 7.7E-08 | 6.9E-09 | 0.2 | 0.3 | 0.6 |
| 316L-XHV; 1 | 1.3E-04 | 0.83 | 2.0E-07 | 5.2E-08 | 7.7E-09 | 0.1 | 0.2 | 0.6 |
| 316L-XHV; 2 | 5.3E-04 | 0.99 | 2.6E-07 | 5.3E-08 | 5.4E-09 | 0.1 | 0.2 | 0.5 |
| 316LN | 4.1E-03 | 1.2 | 5.2E-07 | 7.8E-08 | 5.2E-09 | 0.2 | 0.3 | 0.4 |
| 316LN-XHV | 7.4E-03 | 1.3 | 4.2E-07 | 5.4E-08 | 2.8E-09 | 0.2 | 0.2 | 0.2 |
| Al | 4.2E-03 | 1.0 | 1.6E-06 | 3.2E-07 | 3.0E-08 | 0.7 | 1.2 | 2.5 |
| Ti | 1.3E-03 | 1.0 | 6.4E-07 | 1.3E-07 | 1.3E-08 | 0.3 | 0.5 | 1.1 |





FIGURE CAPTIONS

Figures 5 and 7 are to be 6.75" wide. All others are to be 3.375" wide. Note that the figures embedded in the above text are not the same as the publication quality figures in separate files.

FIG. 1(a) Photograph of all sample chambers used in this study. These are (A) 304L, (B) 316L, (C) 316L-XHV, (D) 316LN, (E) 316LN-XHV, (F) Al, and (G) Ti. (b) Cross-section model view of the sample chamber.

FIG. 2 Schematic of the apparatus used to determine the outgassing rates for the sample chambers. The water outgassing rates are determined using the throughput method and the hydrogen outgassing rates use the rate-of-rise method.

FIG. 3 Solid model showing approximate construction of the apparatus vacuum system. Not depicted is the fore-pump and fore-line that attaches to the back of the turbomolecular pump (Turbo). SRG is the spinning rotor gauge; IG is an ionization gauge; VLV is the all-metal valve; TEE is a union tee; ZRL is a zero-length reducer flange; RTEE is a reducer tee; and RGA is a residual gas analyzer.

FIG. 4 An example of the rate-of-rise data used to determine the specific outgassing rate of the 304L chamber, showing the rise in pressure $p$ in the chamber as a function of time $t$.

FIG. 5 The measured specific outgassing rates for water, $Q_{H2O}$, for all seven sample chambers. (a) The present results for 304L compared to the results of Li and Dylla[1] for chambers vented with air; (b) the present results for 304L compared to those of 316L, 316LN, 316L-XHV, and 316LN-XHV; (c) The present results for 304L compared to those of Ti and Al.

FIG. 6 Repeat measurements of the specific outgassing rates for water, $Q_{H2O}$, for 316L and 316L-XHV. The chambers were removed from the apparatus and left in the ambient





atmosphere between repeat measurements. 316L;1 is the same data presented in Fig. 5(b); 316L;2 is a repeat measurement taken more than 1 year later. 316L-XHV;1 is the same data as in Fig. 5(b); 316L-XHV;2 is a repeat measurement take approximately 9 months later.

FIG. 7 Arrhenius plots of the measured specific outgassing rate for $H_2$ $Q_{H2}$ for six chambers.

FIG. 8 Plots of the measured specific outgassing rate for $H_2$ $Q_{H2}$ for Ti as a function of relative time in days.





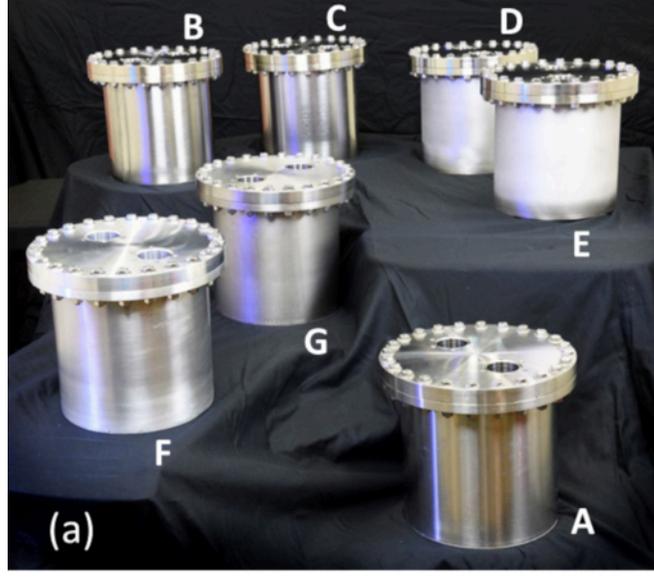

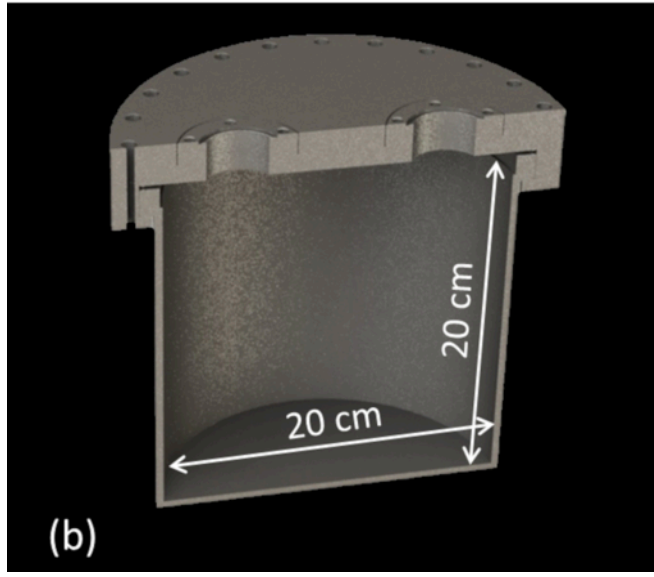



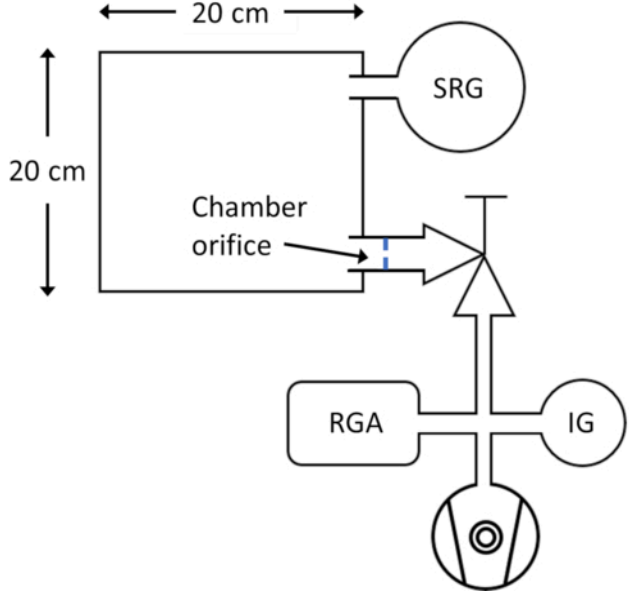



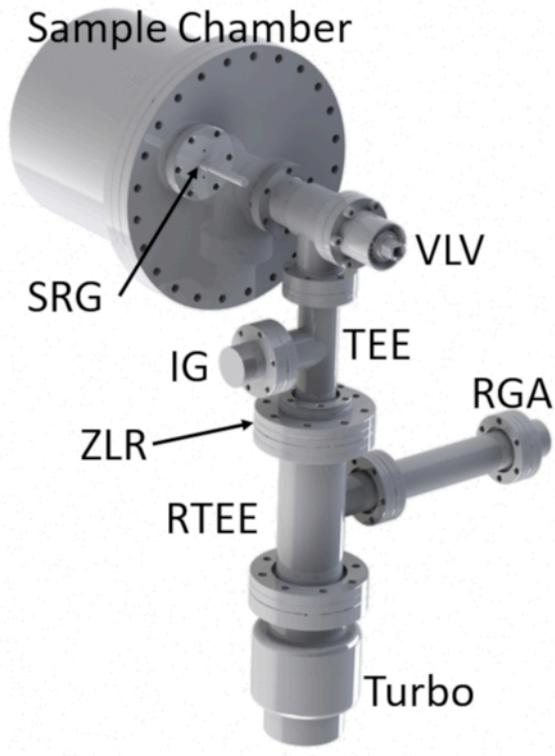

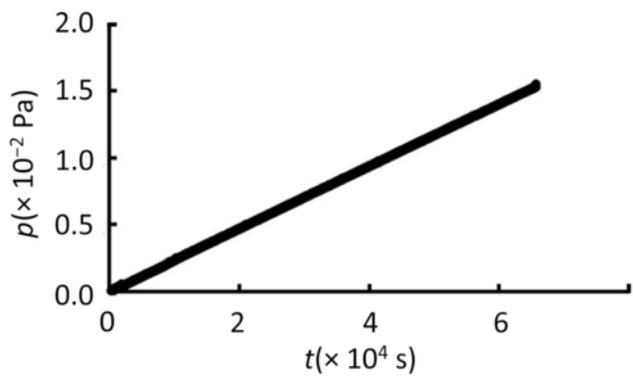



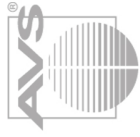



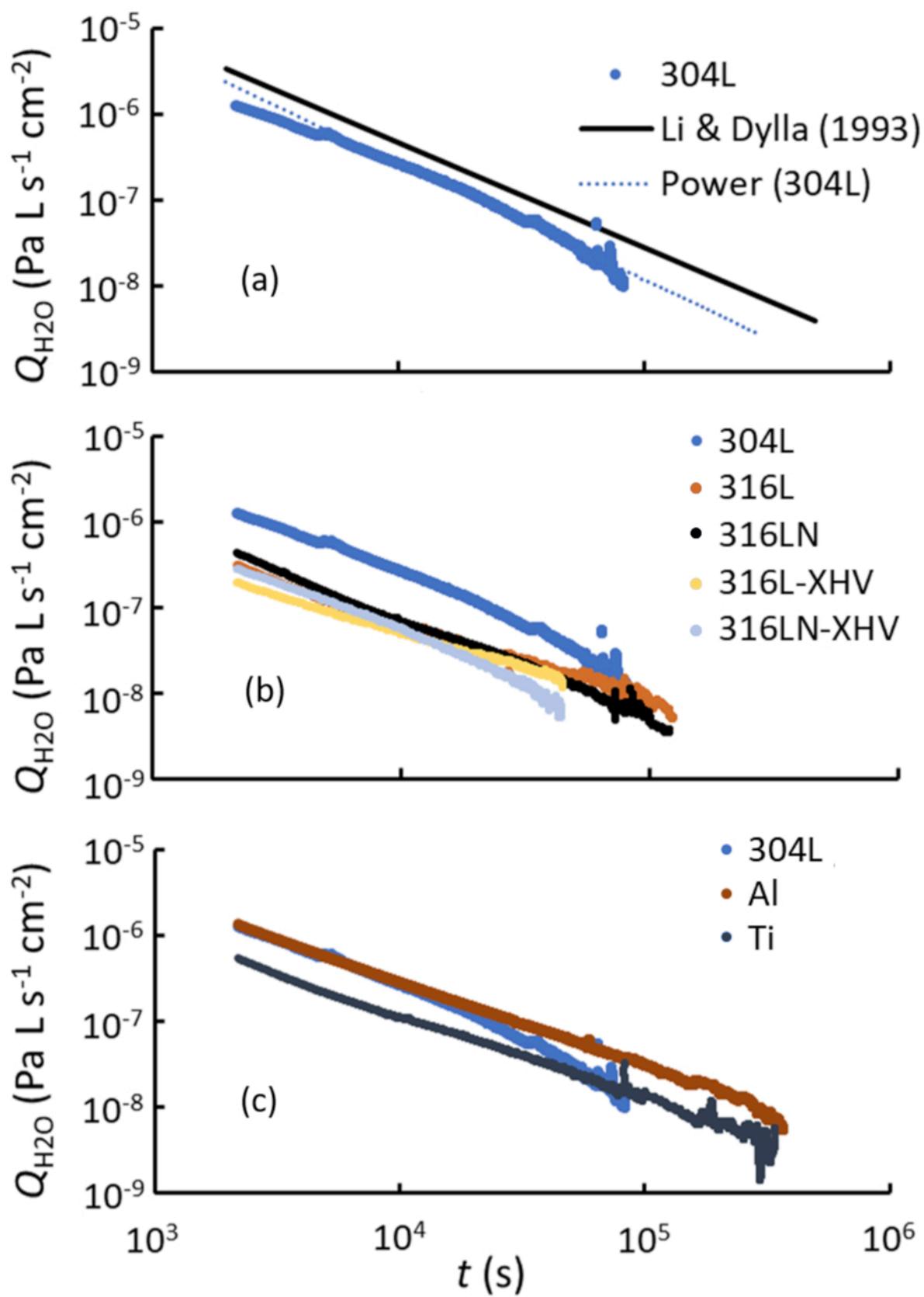



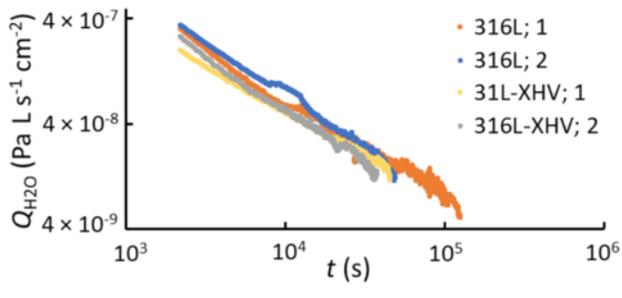

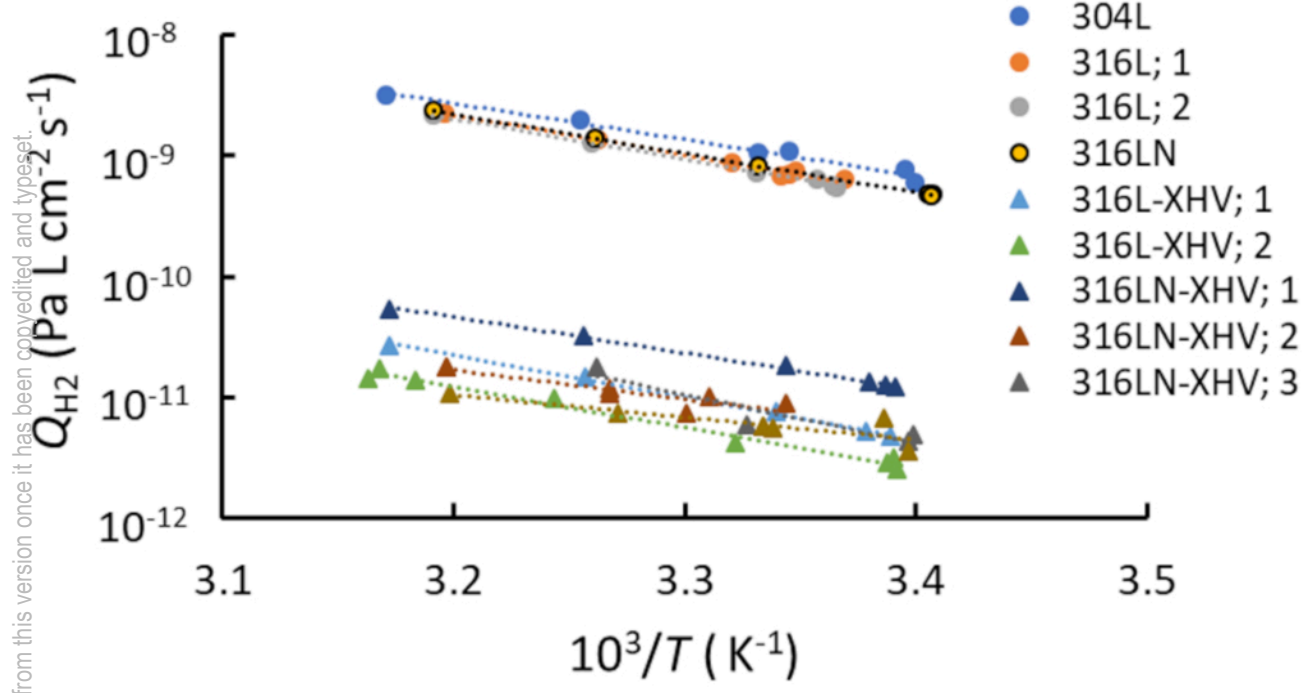







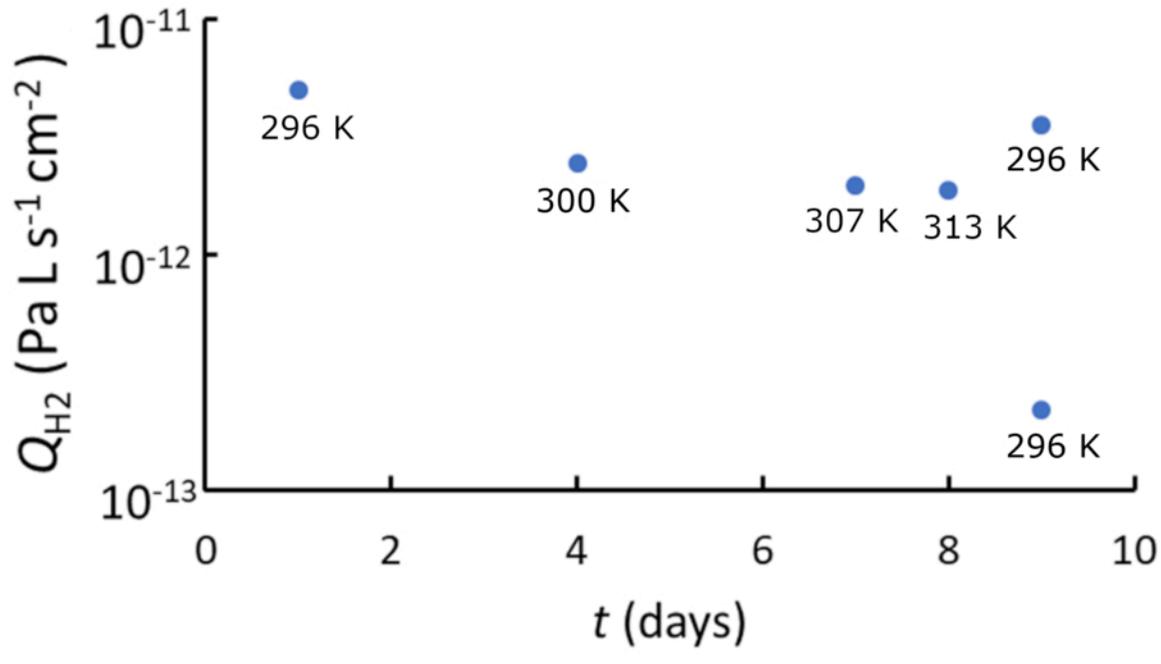